\documentclass[prl,showpacs,twocolumn,preprintnumbers,psfig,amsmath,amssymb]{revtex4}
\usepackage{amsmath}
\usepackage{amsfonts}
\usepackage{amssymb}

\usepackage{epsfig}
\usepackage{graphicx}
\usepackage{dcolumn}
\usepackage{bm}

\newcommand{\beq}{\begin{equation}}
\newcommand{\eeq}{\end{equation}}
\newcommand{\beqa}{\begin{eqnarray}}
\newcommand{\eeqa}{\end{eqnarray}}

\def\gapp{\lower.35em\hbox{$\stackrel{\textstyle>}{\sim}$}}
\def\lapp{\lower.35em\hbox{$\stackrel{\textstyle<}{\sim}$}}

\begin{document}
\bibliographystyle{apsrev}

\title{Gauge fields at the surface of topological insulators}
\author{M. I. Katsnelson}
\affiliation{Radboud University Nijmegen, Institute for Molecules
and Materials, Heyendaalseweg 135, NL-6525 AJ Nijmegen, The
Netherlands}
\author{F. Guinea}
\affiliation{Instituto de Ciencia de Materiales de Madrid (CSIC),
Cantoblanco, Madrid 28049, Spain}
\author{M. A. H. Vozmediano}
\affiliation{Instituto de Ciencia de Materiales de Madrid (CSIC),
Cantoblanco, Madrid 28049, Spain}

\date{\today}
\begin{abstract}
We study the emergence of gauge couplings in the surface states of
topological
insulators. We show that gauge fields arise when a three dimensional strong
topological insulator is coated with an easy--plane ferromagnet with
magnetization parallel  to the surface. We analyze the modification
induced by the gauge fields on  the surface
spectrum for some specific magnetic configurations.

\end{abstract}
%
%
%
%

\maketitle

\section{Introduction}

The Dirac fermion description of novel condensed matter systems
have given rise to a great deal of new phenomena and ideas in
the field that is acting as a bridge between different areas of
physics including quantum field theory, cosmology, elasticity,
and statistical mechanics.

One of the most successful examples can be found in the physics
of graphene \cite{RMP08} and, in particular, in the attractive relation between
the  structural - lattice - properties  and the electronics of this
material. It is by now well known that elastic deformations and
curvature of the graphene lattice give rise to effective gauge
fields that couple to the fermionic degrees of freedom with the minimal
coupling typical of gauge theories
\cite{VKG10}.

Another important example  of Dirac fermions
is that of topological insulators. These materials
are characterized by a bulk insulating
behavior with conducting boundary states protected by some
topological property, the integer quantum Hall effect
\cite{KDP80,AFS82} being the prototype. In the recent developments
(for a review, see Refs. \onlinecite{M09,QZ10,HK10}) numerous systems
have been found where the edge states exist in the absence of a
magnetic field and are protected by time reversal symmetry. The
simplest example is the spin Hall effect \cite{KM05a,BZ06} where
the spin orbit coupling plays a major role;  most of the newly
discovered topological insulators are based on materials with
strong spin orbit coupling \cite{BHZ06,Xetal09,BLetal11}. The topological
protection of edge states ensures their stability against
non-magnetic  -- more generally,  time-reversal invariant --
perturbations, opening up the possibility for future
spintronic devices.

Novel exotic states appear at the surface of a three dimensional
(3D) topological insulator when an energy gap is induced by a
perturbation breaking the time reversal symmetry. They are
obtained typically by applying a magnetic field perpendicular to
the surface what gives rise to the half integer quantum Hall
effect \cite{FK07} or by proximity effects with a magnetic
material what induces an anomalous quantum Hall effect \cite{NSetal10}.
A very recent experiment reports the observation of massive edge
Dirac fermions in a three dimensional topological insulator doped
with magnetic impurities \cite{Cetal10}. In these cases the exotic
physics is due to a combination of the regular or space-dependent
mass induced by the perpendicular magnetic field to the Dirac
fermions,  and by the appearance of zero modes associated to the
Dirac operator in  magnetic fields
\cite{JR76,AC79,JR81}.

A different situation arises when the surface of the topological
insulator is coated with an insulating ferromagnet with an easy
plane such that the magnetization is parallel to the surface.
Although time reversal invariance is also broken and edge states
are no more protected, a mass is not directly induced in the
system. In this work we will see that in such situation
gauge  fields appear
coupling to the electronic degrees of freedom similar to these
produced by elastic deformation on graphene. We will explore the
modification of the spectrum for the cases of a vortex and a skyrmion
configuration.

\section{Gauge field induced by an in-plane magnetization}

When putting a thin ferromagnetic layer with magnetization ${\vec
M}$ on the surface of a strong topological insulator
an interaction term  of the form
$H'=-J {\vec M}.{\vec \sigma}$  arises where $J$
is the coupling constant.  There are two contributions to this
magnetic term: the Zeeman energy, and $s-d$ exchange interaction
energy. The latter contribution can be orders of magnitude higher
than the former one if the overlap between the electron
 wave function on
the surface of the topological insulator and that of the
ferromagnet is big enough. One can easily expect that $JM\sim$
0.01-0.1 eV. This term breaks the time reversal symmetry. In the
case of an easy-axis ferromagnet with the magnetization perpendicular
to the layer the magnetic interaction has a structure of a mass term proportional
to $\sigma_z$ in the
Dirac Hamiltonian resulting in the gap opening mentioned above.
In the case of having an easy-{\it plane} ferromagnet, the vector
${\vec M}$ lies in the surface plane and the $s-d$ exchange
interaction couples to the fermions as a vector potential.

In the case of graphene where the ``spin'' degree of freedom
is actually a {\it pseudo}spin related
to the sublattice degree of freedom, gauge couplings are induced by
strains or other mechanical deformations of the
lattice \cite{VKG10}. In topological insulators where spin is the
real spin one needs in-plane magnetization to create such terms.

Easy plane ferromagnetism is very natural for thin films of single
layers as the shape anisotropy (magnetic dipole--dipole
interaction)  always favor this situation \cite{V74}. This will be
the case for any soft magnetic material i. e. materials fulfilling
the condition $4\pi M_s^2>K$, $K$ being the constant of
magnetocrystalline anisotropy. If the coating is a circular
symmetric disc, the magnetization vector will be freely rotating
in the plane.  Magnetic textures at the surface of a three
dimensional topological insulator have  been recently studied in
Refs. \onlinecite{NN10,ZZ10}. In the latter publication, the
exchange field energy associated to parallel ferromagnetic domains
of 50 nm width was estimated to be of the order of 6 meV.

\subsection{Vortex configuration}
\begin{figure}
\begin{center}
\includegraphics[height=4cm]{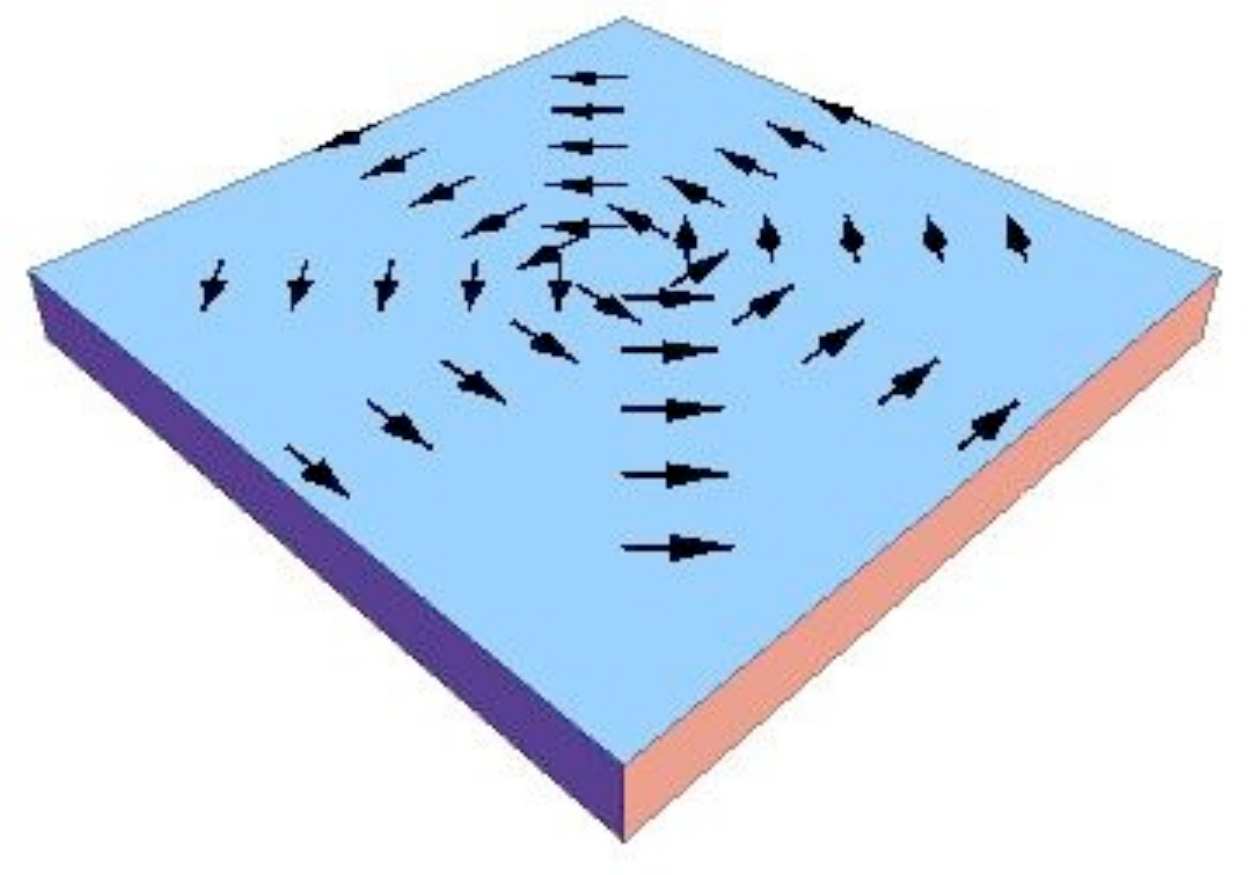}
\caption{(Color online) Sketch of the ferromagnetic vortex configuration. }
\label{vortex}
\end{center}
\end{figure}
In easy--plane ferromagnets vortices can exist which are spin
configurations with a net $2\pi$ twist about a particular point or
core similar to the Onsager-Feynman vortices in superfluid helium
or Abrikosov vortices in superconductors. Examples are BaCoAsO$_4$
and K$_2$CuF$_4$ \cite{RR74}.

In this case the magnetization is constant in magnitude and
rotates in direction  as shown schematically in Fig. \ref{vortex}.
The corresponding vector potential in polar coordinates is $A_r=0,
A_\theta=-JM$.

Let us assume, for simplicity, that without interaction with
magnetization the electron spectrum corresponds to that of
isotropic (in plane) Dirac cone. Then, the dynamics of the low
energy states of the system in the presence of the vortex is
described in polar coordinates by the Hamiltonian

\beq
{\cal H} =  i\left( \begin{array}{cc} 0 &
 e^{-i\theta}\left(\partial_r - i \frac{\partial_\theta}{r} + JM \right)
 \\ e^{i\theta}\left(\partial_r + i \frac{\partial_\theta}{r} - JM  \right ) & 0
 \end{array} \right). \label{hamvortex}
\eeq where we have put $\hbar=1$. From now on and since we will
not deal with interactions which can renormalize the Fermi
velocity \cite{GGV94} we will put it just to unity: $v=1$.

The usual ansatz
\beq
\Psi_1(r,\theta)=e^{il\theta}\varphi_1(r)
\;\;\;,\;\;\; \Psi_2(r,\theta)=e^{i(l+1)\theta}\varphi_2(r),
\eeq
allows to write the Schrodinger equation $H\Psi=E\Psi $ with
$\Psi=(\Psi_1,\Psi_2)^{\dagger}$, as \beqa
i\left(\frac{d}{dr}+\frac{l+1}{r}+JM\right)\varphi_2 & =E\varphi_1
\\\nonumber \vspace{0.3cm}
i\left(\frac{d}{dr}-\frac{l}{r}-JM\right)\varphi_1 & =E\varphi_2,
\eeqa which can be reduced to
\beq
\left[\frac{d^2}{d r^2}+\frac{l}{r}\frac{d}{d r}-
\frac{l(l+1)}{r^2}+\frac{JM(2l+1)}{r}\right]\varphi
=\left[(JM)^2-E^2\right]\varphi.
\label{hydrogen}
\eeq
for both components.

Equation (\ref{hydrogen}) is formally identical to the
Schr\"{o}dinger equation of a planar hydrogen atom with a nucleus
of charge $Ze$ \cite{YGetal91a,KH98}:
\beq \left(\Delta-2 m \frac{Ze^2}{r}\right)\varphi=-2m\epsilon \varphi ,
\eeq
with the substitution $\varepsilon=E^2-(JM)^2$, $2mZe^2=JM(2l+1)$.
From this formal analogy we can write the answer for the
eigenvalues of the energy:
\beq
E^2(n_r,l)=(JM)^2\left[1-\frac{(l+1/2)^2}{(n_r+l+1/2)^2}\right],
\label{specvortex}
\eeq with $n_r=0, 1, 2..$.

\begin{figure}
\begin{center}
\includegraphics*[width=6cm]{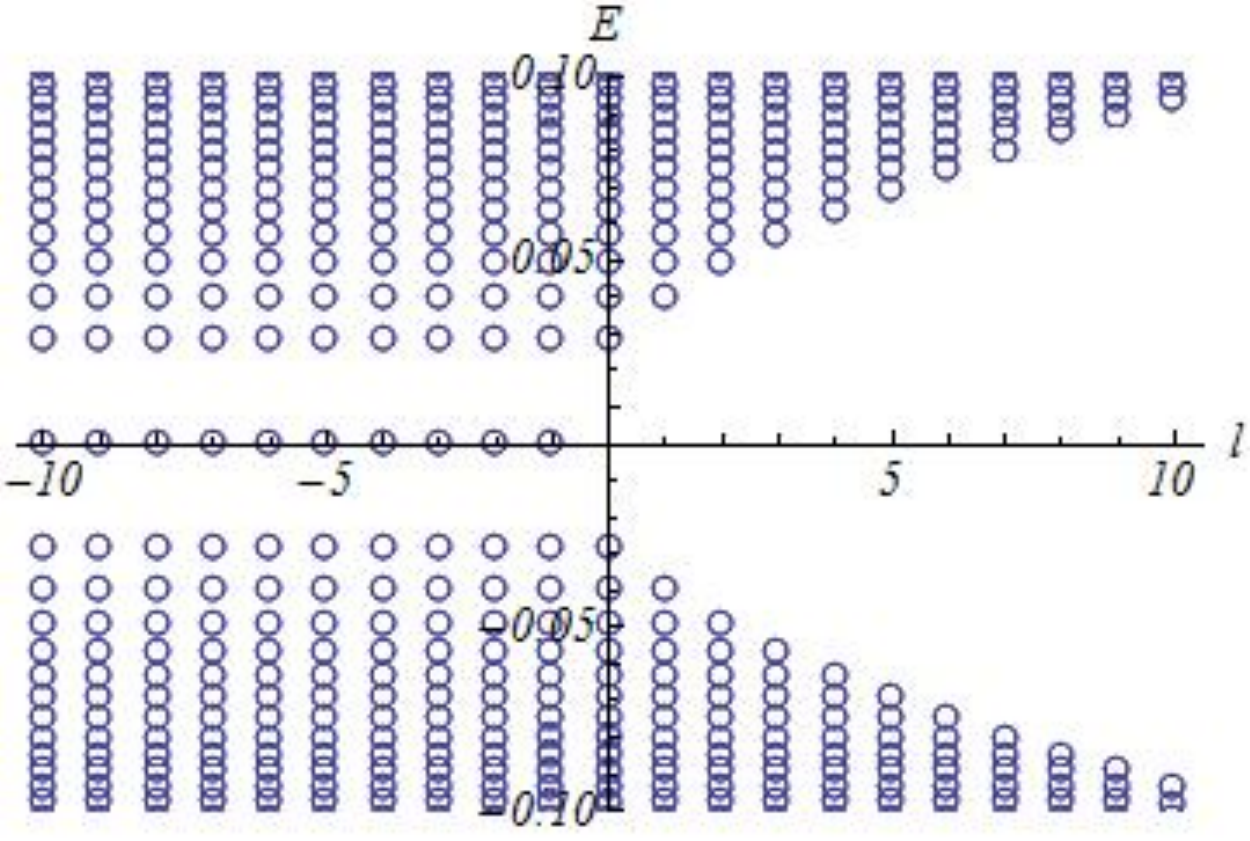}
\includegraphics*[width=6cm]{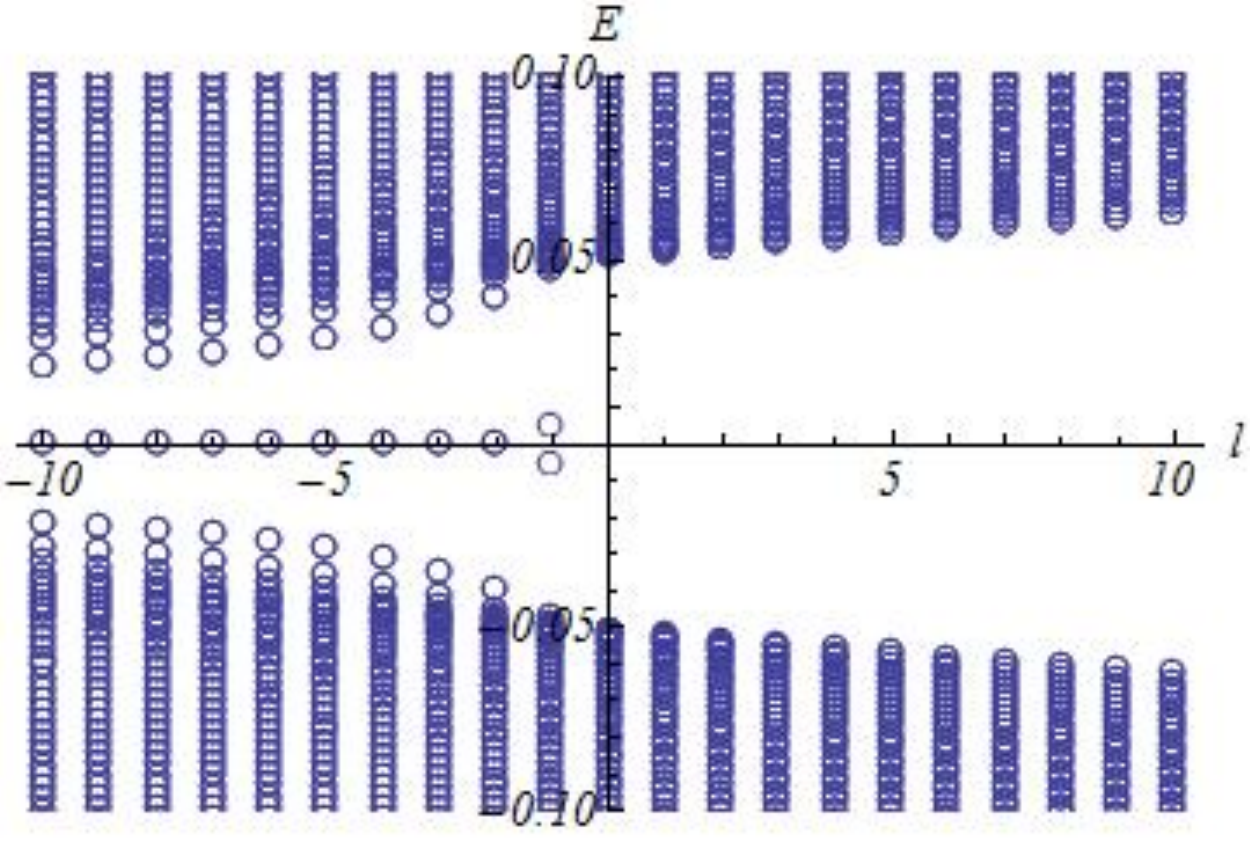}
\includegraphics*[width=6cm]{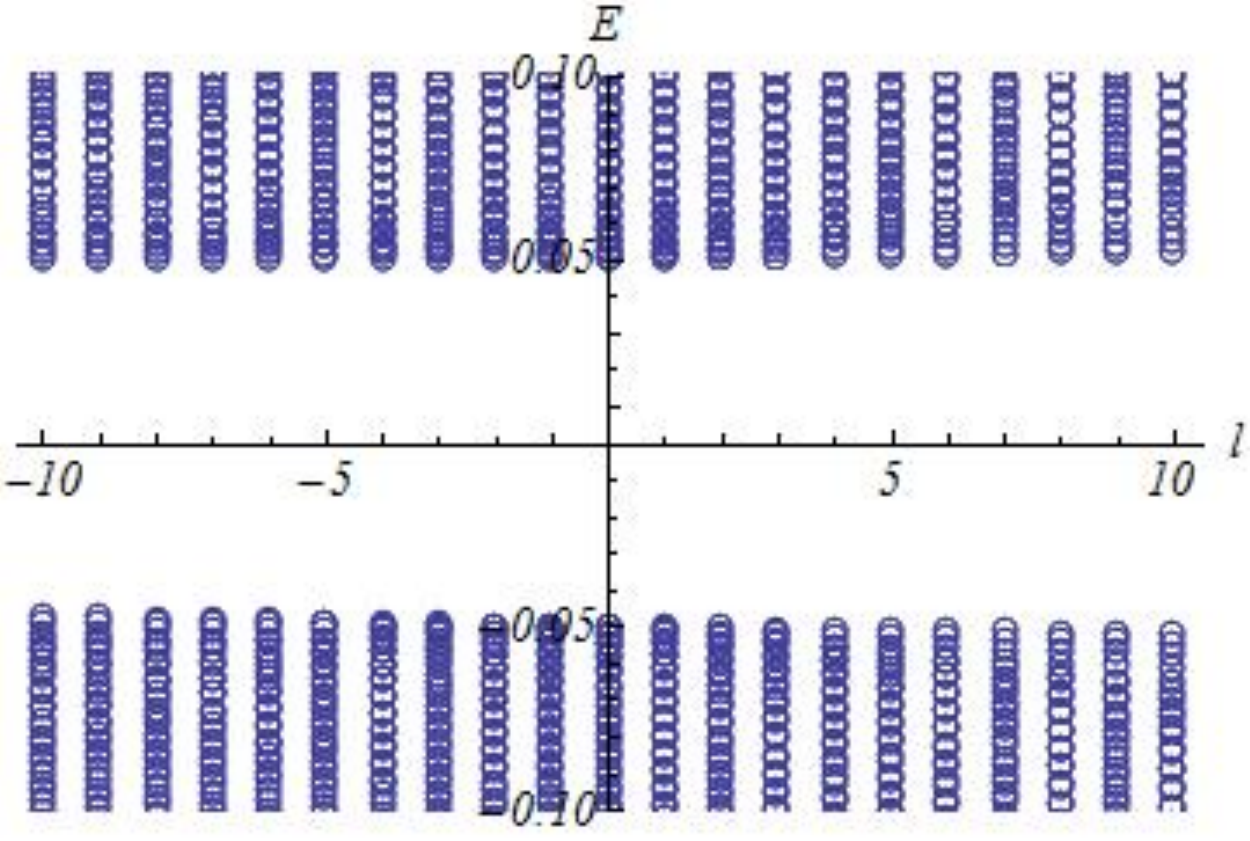}
\caption{Top: Energy levels as function of angular momentum at the surface of a topological insulator in a constant in--plane magnetic field. Center: Energy levels in the presence of a magnetic vortex. Bottom: Energy levels in the presence of a skyrmion configuration.}
\label{figvortex}
\end{center}
\end{figure}

To ease the readout of the spectrum we show a plot of the numerical
solution of eq. (\ref{hamvortex}) in Fig. \ref{figvortex}
(center). The results reproduce  the analytical spectrum given in eq.
(\ref{specvortex}) but only for one sign of the angular momentum.
There are no bound states for the other sign. For comparison we
also plot the numerical results for the spectrum of a  disc as a
function of the angular momentum in the presence of a constant magnetic
field (Fig. \ref{figvortex} (a)).

We can see that for $2l+1 >0$ (assuming $JM>0$)
there are  localized states in the energy interval  $\vert E\vert<JM$ with
accumulation points at the boundary values  $E=\pm JM$.
The numerical solution shows also that
the wave functions for the localized states decay
exponentially at large distances. For any value of  $l$ there is a
zero-energy mode with $n_r =0$.

For $2l+1<0$ we have the effective Schr\"{o}dinger equation for
the Coulomb center with repulsion, and no localized states.

The natural spatial scale in the problem is $\xi=\frac{\hbar
v_F}{\vert J\vert M}$ which plays the same role as the correlation
length for the Abrikosov vortices.

\subsection{Skyrmion configuration}
A novel situation arises when the (3D) magnetic coating is in a
skyrmion configuration \cite{S61}. 2D skyrmion patterns can exist when the magnetic anisotropy is weak, and they have been recently
observed with Lorentz transmission electron microscopy in a thin
film of Fe$_{0.5}$Co$_{0.5}$Si \cite{YOetal10}. The magnetization
vector evolves from  pointing along a given direction at the center of the defect to the opposite direction at large distances. The absolute magnetization does  not change, and at intermediate distances it has an in-plane component. A possible skyrmion texture is
\begin{align}
{\bf M}(r, \theta) &= (\sin f(r)\cos\theta,
\sin f(r)\sin\theta, \cos f(r) ),
\end{align}
where $f(r)$ can be chosen as a constant for simplicity.

A numerical solution of the  spectrum induced by a skyrmion texture  is shown in  Fig. \ref{figvortex} (bottom). The calculation solves an effective one dimensional differential equation for each value of the angular momentum\cite{WSG08}. The figure also shows spectra for a constant magnetic field (top), and for the gauge field associated to a vortex (center). The skyrmion opens a gap in the spectrum and does not support a zero energy solution.
Note that a zero mode appears when the in-plane component of the magnetization is subtracted.

\section{Conclusions and discussion}

This work originated on the question of whether an effective gauge
field would be induced by strain on the surface of a topological
insulator. As it is known this is the case in graphene which can
be seen in many respects of a precursor of the actual topological
insulators. It is known that elastic deformations and geometrical
corrugations in graphene can be described by and induced
fictitious gauge field coupling to the electronic degrees of
freedom \cite{VKG10}. The apparent paradox that the strains do not
break time reversal symmetry while magnetic fields do is solved in
the graphene case by the fact that the induced fictitious field
couples to the two Dirac cones with opposite signs and time
reversal symmetry  is preserved in the complete system. This
mechanism will probably occur also in the case of weak topological
insulators with an even number of Dirac cones at the surface. The
response of a strong topological insulator to elastic deformations
has to be different and remains to be studied. In this work we
have seen an alternative way to generate gauge fields in a strong
topological insulator and we have studied the modification of the
energy spectrum caused by the gauge fields.

Magnetic coating of the surface three dimensional TI is often
invoked to describe specific physical situations where a  gap
is needed in the 2D system \cite{Q09,J09,GC11}. The discussion
presented in this work shows that in specific situations
the magnetic coating coating can
give rise to  a strong reorganization
of the spectrum that has to be taken into account for
each physical situation.

\section{Acknowledgments}
\label{thank}

MAHV thanks A. Cortijo and A.G.  Grushin for
discussions. FG and MAHV acknowledge support from MEC (Spain) through
grant FIS2005-05478-C02-01, PIB2010BZ-00512 and CONSOLIDER CSD2007-00010, and by
the Comunidad de Madrid, through CITECNOMIK,
CM2006-S-0505-ESP-0337. MK acknowledges support from Stichting
voor Fundamenteel Onderzoek der Materie (FOM), the Netherlands.

\bibliography{Topo}
\end{document}